\documentclass[copyright]{eptcs}
\providecommand{\event}{PAR-2010 Edinburgh, UK} 
\usepackage {bsymb,b2latex,amsmath,proof,listings, graphicx}
\graphicspath{{./Figures/}}   
\newtheorem{theorem}{Theorem}[section]
\newtheorem{lemma}[theorem]{Lemma}
\newtheorem{proposition}[theorem]{Proposition}

\newtheorem{definition}[theorem]{Definition}
\title{Rewriting and Well-Definedness within a Proof System\thanks{This research is carried out as part of DEPLOY (an European Commission FP7 Project Grant 214158). The first author is partially supported by the Algerian Ministry of Higher Education (MESRS).}}
\author{Issam Maamria and Michael Butler
\email{\{im06r,mjb\}@ecs.soton.ac.uk}
\institute{Electronics and Computer Science\\}
\institute{University of Southampton\\ Southampton, UK}
}
\def\titlerunning{Rewriting and Well-Definedness within a Proof System}
\def\authorrunning{Issam Maamria and Michael Butler}
\begin{document}
\maketitle
\begin{abstract}
Term rewriting has a significant presence in various areas, not least in automated theorem proving where it is used as a proof technique. Many theorem provers employ specialised proof tactics for rewriting. This results in an interleaving between deduction and computation (i.e., rewriting) steps. If the logic of reasoning supports partial functions, it is necessary that rewriting copes with potentially ill-defined terms. In this paper, we provide a basis for integrating rewriting with a deductive proof system that deals with well-definedness. The definitions and theorems presented in this paper are the theoretical foundations for an extensible rewriting-based prover that has been implemented for the set theoretical formalism Event-B.
\end{abstract}

\section{Introduction}
Term rewriting has an important presence in many areas including abstract data type specifications and automated reasoning. In this regard, many automated theorem provers employ rewriting as a proof technique where it may interleave with deduction. PVS~\cite{886667} and Isabelle/HOL~\cite{513715} are higher-order theorem provers that include specialised tactics for rewriting.
\par
The interleaving between rewriting steps and deduction steps poses several difficulties. The termination of rewriting becomes an issue of paramount importance. Many techniques, such as term orderings~\cite{280474}, have been explored to provide good practical solutions to termination problems. We argue that, in the presence of potentially ill-defined terms, rewriting has to be further constrained.
\par
Ill-defined terms arise in the presence of partial functions. They result from the application of functions to terms outside their domain. If ill-definedness is a concern, the adopted reasoning framework has to cope with it. Different approaches exist to reason in the presence of partial functions. Each of these approaches has its own specialised proof calculus. In~\cite{icfemMehta08}, it is shown that it is possible to reason about partiality without abandoning the well-understood domain of two-valued predicate logic. In that approach, the reasoning is achieved by extending the standard calculus with derived proof rules that preserve well-definedness across proofs. We argue that, in order to integrate rewriting as a proof step in such a calculus, it is necessary that rewriting preserves well-definedness.
\par
In this paper, we present a treatment of term rewriting where term well-definedness is an issue. Our treatment unifies the notions of well-definedness (WD) and rewriting, and provides a basis to integrate rewriting as a proof step within the proof system presented in~\cite{icfemMehta08}. Central to our contribution is the concept of WD-preserving rewriting where rewrite rules preserve well-definedness in the same direction in which they are applied. We establish the necessary conditions under which rewriting preserves well-definedness. We, finally, show how a rewrite step can be interleaved with deduction steps in a valid fashion.

\subsection{Practical Setting}\label{pracSetting}
Event-B~\cite{1365974} is a formalism for discrete systems modelling based on Action Systems~\cite{91938}. It can be used to model and reason about complex systems such as concurrent and reactive systems. The semantics of a model developed in Event-B is given by means of its proof obligations. These obligations have to be discharged to show consistency of the model with respect to some behavioural semantics.
\par
\begin{figure}[tbp]
\includegraphics[width=4.5in, height=5in]{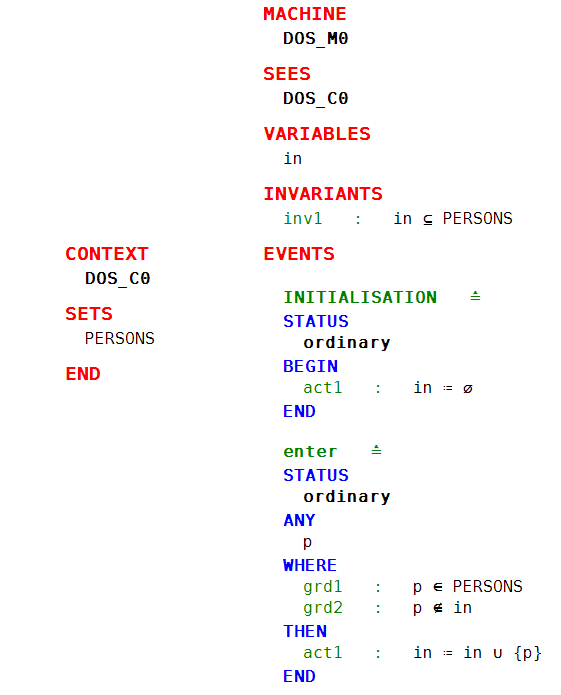}
\caption{A Simple Model for A Door Entry System}\label{simpleModel}
\end{figure}

Modelling in Event-B is conducted by defining contexts and machines. Contexts describe static properties of a model by specifying carrier sets and constants. Machines, as their name suggests, define the dynamics of a model by means of variables (state) and events (transitions). Variables are constrained by invariants. A machine can be refined by another machine, and can see (import) contexts. Proof obligations arise to verify the consistency of a model. For instance, there are proof obligations to establish the refinement relationship between two machines, and to establish invariant preservation by the events (transitions). The logic used in Event-B is typed set theory built on first-order predicate logic, and allows the definition of partial functions. As such, it is necessary that the used proof system handles ill-definedness. Indeed, the proof calculus outlined in~\cite{icfemMehta08} is the one used to reason in Event-B. Figure \ref{simpleModel} illustrates a simple Event-B model for a door entry system.
\par
The Rodin platform~\cite{Abrial-etal06} is an open extensible tool for Event-B based on Eclipse\footnote{http://www.eclipse.org/}. It offers support
for specification and proof, and it can be easily extended with other useful tools e.g., there is a plug-in for model checking called Pro-B~\cite{LeuschelButler:FME03}.
\subsection{Motivation}
The Rodin platform has a proving infrastructure which is extensible with new proof rules. External provers can also be used; Atelier-B~\cite{abrial_03_clickn} provers ML and PP have been incorporated into Rodin. Adding new proof rules requires the use of the Java programming language, knowledge of Eclipse as well as an understanding of the internal architecture of Rodin. A complication of such approach is that newly implemented rules could compromise the soundness of the prover. This work has been carried out as part of an effort to address this limitation of Rodin from the viewpoint of prover extensibility. This paper discusses some theoretical results in the context of rewriting and well-definedness. The ideas presented in this paper have resulted in providing proof support for the set theoretical formalism Event-B~\cite{1365974}. An extensible rewriting-based prover~\cite{issam1984} has been implemented and integrated into Rodin. 
\par
\textit{Outline.} In Section \ref{pre}, we recall some preliminary concepts of term rewriting systems. Section \ref{wdpres} describes the necessary conditions under which rewriting preserves well-definedness. Section \ref{pstep} shows how a WD-preserving rewrite rule can be used in proofs. The application of the previous ideas in the context of Event-B~\cite{1365974} is shown in Section \ref{eapp}. We conclude in Section \ref{future} by stating what we have achieved and its impact on the Event-B toolset~\cite{3056425}.
\subsection{Related Work}
The interleaving between deduction and rewriting steps has gathered much interest given its importance to automated reasoning. In this work, we identify the necessary conditions under which rewriting can interleave with deduction in the proof calculus defined in~\cite{icfemMehta08}. In other works, this interleaving is studied from different perspectives.
\par
Theorem proving modulo is an approach that removes computational steps from proofs by reasoning modulo a congruence on propositions~\cite{949773}. The advantage of this technique is that it separates computation steps (i.e., rewriting) from deduction steps in a clean way. In~\cite{949773}, a proof-theoretic account of the combination between computations and deductions is presented in the shape of a sequent calculus modulo. The congruence on propositions, on the other hand, is defined by rewrite rules and equational axioms.
\par
The combination of rewriting and deduction makes properties of rewrite systems of practical interest. Termination and confluence properties of term rewriting systems are important, and have been studied extensively~\cite{280474,303448}. When rewriting is interleaved with deduction, it is critical that computation steps terminate. Term orderings, in which any term that is syntactically simpler than another is smaller than the other, provides a practical technique to assess the termination of rewrite systems.
\par
In our work, we aim to unify the notions of well-definedness and rewrite systems. Our objective is to characterise the interaction between deduction and rewriting when well-definedness is taken into consideration. This is achieved by identifying the necessary conditions under which computations can interleave with the deduction steps (i.e., proof rules) in~\cite{icfemMehta08}.

\section{Preliminaries}\label{pre}
In this section, we lay the groundwork for the rest of the paper. We briefly introduce the proof calculus defined in~\cite{icfemMehta08}. We also shed some light on basic concepts of term rewriting systems. For the rest of this paper, we use the language signature $\Sigma$ defined by a set $V$ of variable symbols, a set $F$ of function symbols and a set $P$ of predicate symbols. In the next two definitions, we introduce the syntax of the first-order predicate calculus with equality that will be used in the subsequent sections.

\begin{definition}[Term]
$T_{\Sigma}$, the set of $\Sigma$-terms is inductively defined by:\\
~~$\bullet$ each variable of $V$ is a term;\\
~~$\bullet$ if $f\in F$, $arity(f) = n$ and each of $e_1,...,e_n$ is a term, then $f(e_1,...,e_n)$ is a term.
\end{definition}
\begin{definition}[Formula]
$F_{\Sigma}$, the set of $\Sigma$-formulas is inductively defined by:\\
~~$\bullet$ $\bot$ is a formula;\\
~~$\bullet$ $p(t_1,...,t_n)$ is a formula provided $p\in P$, $arity(p)=n$ and each of $t_1,...,t_n$ is a term;\\
~~$\bullet$ $t_1 = t_2$ is a formula provided $t_1$ and $t_2$ are terms;\\
~~$\bullet$ $\varphi \land \psi$ is a formula if $\varphi$ and $\psi$ are formulas;\\
~~$\bullet$ $\lnot \varphi$ is a formula if $\varphi$ is a formula;\\
~~$\bullet$ $\forall x.\varphi$ is a formula if $x \in V$ and $\varphi$ is a formula.
\end{definition}
Note that other logical operators (e.g., $\exists$) can be defined (as in~\cite{farhad}) by means of the operators in the previous definition. For the rest of the paper, we assume a syntactic operator $\mathcal{V}ar: (F_{\Sigma} \cup T_{\Sigma})\rightarrow \pow(V)$ such that $\mathcal{V}ar(t)$ is the set of variables occurring free in $t$.

\subsection{The Well-Definedness Operator}
The well-definedness operator '$\mathcal{D}$' encodes what is meant by well-definedness. $\mathcal{D}: (F_{\Sigma}\cup T_{\Sigma})\rightarrow F_{\Sigma}$ is a syntactic operator that maps terms and formulae to their well-definedness predicates. We interpret the formula $\mathcal{D}(F)$ as being valid if and only if $F$ is well-defined.
For a detailed treatment of the $\mathcal{D}$ operator, we refer to~\cite{723108}.
\par
The well-definedness (WD) of terms is defined recursively as follows:
\begin{eqnarray}
\mathcal{D}(x)~&\widehat{=}&~\top~~~if~x\in V~\label{wdVar}\\
\mathcal{D}(f(t_1,...,t_n))~&\widehat{=}&~\bigwedge_{i=1}^n \mathcal{D}(t_i)~\land~C^f_{t_1,...,t_n}~\label{wdFun}.
\end{eqnarray}
where $C^f_{t_1,...,t_n}$ effectively defines the domain of the function $f$. For this study, we assume that predicate symbols are total. As a result, ill-definedness can only be introduced by terms. Therefore, we have the following:
\begin{eqnarray}
\mathcal{D}(p(t_1,...,t_n))~&\widehat{=}&~\bigwedge_{i=1}^n \mathcal{D}(t_i)~~~if~p\in P~\label{wdPred}\\
\mathcal{D}(t_1 = t_2)~&\widehat{=}&~\mathcal{D}(t_1) \land \mathcal{D}(t_2)~\label{wdEq}.
\end{eqnarray}
For the well-definedness of other formulae, we use the following expansions~\cite{723108}:
\begin{eqnarray}
\mathcal{D}(\bot)~&\widehat{=}&~\top\\
\mathcal{D}(\lnot \varphi)~&\widehat{=}&~\mathcal{D}(\varphi)\\
\mathcal{D}(\varphi \land \psi)~&\widehat{=}&~(\mathcal{D}(\varphi)\land \mathcal{D}(\psi))\lor(\mathcal{D}(\varphi) \land \lnot \varphi)\lor(\mathcal{D}(\psi)\land \lnot \psi)\\
\mathcal{D}(\forall x\cdot \varphi)~&\widehat{=}&~(\forall x\cdot \mathcal{D}(\varphi))\lor(\exists x\cdot \mathcal{D}(\varphi)\land \lnot \varphi)
\end{eqnarray}
The well-definedness of formulae built using derived logical operators can be straightforwardly derived, see~\cite{farhad}. An important property of well-definedness conditions is that they are themselves well-defined~\cite{icfemMehta08}; i.e.,
\begin{equation*}
\mathcal{D}(\mathcal{D}(P))~\leqv~\top~.
\end{equation*}

\subsection{The WD-preserving Sequent Calculus}
We assume the signature $\Sigma$ is equipped with a proof theory in the shape of a WD-preserving first-order sequent calculus similar to the one appearing in~\cite{icfemMehta08}. A judgement in the aforementioned calculus is called a well-defined sequent, and is of the form $H\vdash_{_\mathcal{D}} G$ defined as follows:
\begin{eqnarray*}
H\vdash_{_\mathcal{D}} G~~\defi~~ \mathcal{D}(H),\mathcal{D}(G),H\vdash G~.
\end{eqnarray*}
That is, the well-definedness of $H$ and $G$ is assumed when proving $H\vdash G$. Generally speaking, when proving a sequent $H\vdash G$, the approach suggests proving its validity as well as its well-definedness:
\begin{eqnarray*}
\boxed{\text{WD}_{_\mathcal{D}}:~~~\vdash_{_\mathcal{D}} \mathcal{D}(H\vdash G)}~~~~~~~~~~\boxed{\text{Validity}_{_\mathcal{D}}:~~~H \vdash_{_\mathcal{D}} G}
\end{eqnarray*}
where $\mathcal{D}(H\vdash G)$ is defined as $\mathcal{D}(\forall \overrightarrow{x}\cdot H\limp G)$ such that $\overrightarrow{x}$ are the free variables of $H$ and $G$.
\par
A proof rule is said to preserve well-definedness (WD) iff its consequent and antecedents only contain well-defined sequents (i.e., $\vdash_{_\mathcal{D}}$ sequents). Figure \ref{fopced} introduces the theory \textit{\textbf{FoPCe}}$_{_\mathcal{D}}$ (a collection of WD-preserving inference rules) as developed in~\cite{icfemMehta08}. Note that we use $x\backslash H$ to denote the non-freeness condition of $x$ in $H$.  We also use $[x:=E]P$ to denote the syntactic replacement of all free occurrences of the variable $x$ in $P$ by the term $E$. The boxed sequents in Figure \ref{fopced} correspond to the additional sequents that has to be discharged compared to the classical version of the rule in order to preserve well-definedness.
\begin{figure}[h]
\fbox{
\begin{minipage}{15.5cm}
\begin{center}
$
\infer[hyp_{_\mathcal{D}}]{H, P\vdash_{_\mathcal{D}} P}{}~~~~~~~~~~\infer[mon_{_\mathcal{D}}]{H, P\vdash_{_\mathcal{D}} Q}{H \vdash_{_\mathcal{D}} Q}~~~~~~~~~~
\infer[contr_{_\mathcal{D}}]{H\vdash_{_\mathcal{D}} Q}{H, \lnot Q\vdash_{_\mathcal{D}} \bot}
$
\end{center}
\begin{center}
$
\infer[\bot hyp_{_\mathcal{D}}]{H,\bot \vdash_{_\mathcal{D}} P}{}~~~~~~~~~~\infer[\lnot goal_{_\mathcal{D}}]{H\vdash_{_\mathcal{D}}\lnot P}{H, P\vdash_{_\mathcal{D}}\bot}~~~~~~~~~~\infer[\lnot hyp_{_\mathcal{D}}]{H,\lnot P\vdash_{_\mathcal{D}} Q}{H \vdash_{_\mathcal{D}} P}
$
\end{center}
\begin{center}
$
\infer[\land goal_{_\mathcal{D}}]{H\vdash_{_\mathcal{D}} P\land Q}{H\vdash_{_\mathcal{D}}P~~H\vdash_{_\mathcal{D}}Q}~~~~~~~~~~
\infer[\land hyp_{_\mathcal{D}}]{H, P\land Q \vdash_{_\mathcal{D}} R}{H, P, Q \vdash_{_\mathcal{D}} R}~~~~~~~~~~
\infer[\forall goal_{_\mathcal{D}}~(x\backslash H)]{H \vdash_{_\mathcal{D}}\forall x \cdot P}{H \vdash_{_\mathcal{D}} P}
$
\end{center}
\begin{center}
$
\infer[=goal_{_\mathcal{D}}]{H\vdash_{_\mathcal{D}} E=E}{}~~~~~~~~~~
\infer[=hyp_{_\mathcal{D}}]{H, E=F\vdash_{_\mathcal{D}} [x:=F]P}{H\vdash_{_\mathcal{D}} [x:=E]P}
$
\end{center}
\begin{center}
$
\infer[cut_{_\mathcal{D}}]{H\vdash_{_\mathcal{D}} Q}{\boxed{H \vdash_{_\mathcal{D}} \mathcal{D}(P)}~~~~~~~~~~H \vdash_{_\mathcal{D}} P~~~~~~~~~~H,P\vdash_{_\mathcal{D}} Q}
$
\end{center}
\begin{center}
$
\infer[\forall hyp_{_\mathcal{D}}]{H,\forall x \cdot P\vdash_{_\mathcal{D}} Q}{\boxed{H \vdash_{_\mathcal{D}} \mathcal{D}(E)}~~~~~~~~~~H,[x:=E]P\vdash_{_\mathcal{D}} Q}
$
\end{center}
\end{minipage}
}
\caption{Inference Rules of \textbf{\textit{FoPCe}}$_{_\mathcal{D}}$}
\label{fopced}
\end{figure}
\newline Proof rules for derived logical operator (i.e., $\limp$, $\lor$, $\leqv$ and $\exists$) can be derived directly from the rules of \textbf{\textit{FoPCe}}$_{_\mathcal{D}}$. The following two proof rules can be derived with a detour through $\vdash$ sequents (classical reasoning):
$$
\infer[goal_{_{WD}}]{P~\vdash_{_\mathcal{D}}~R}{P, \mathcal{D}(R)~\vdash_{_\mathcal{D}}~R}
$$
and
$$
\infer[hyp_{_{WD}}]{P~\vdash_{_\mathcal{D}}~R}{P, \mathcal{D}(P)~\vdash_{_\mathcal{D}}~R}~.
$$
\par
In Section \ref{wdpres} and \ref{pstep}, we show how rewriting can be interleaved with the inference rules of \textbf{\textit{FoPCe}}$_{_\mathcal{D}}$. For the rest of the paper, we assume that the reader is familiar with the basic notions of rewriting as found, for instance, in~\cite{280474}. We define the domain and range of a substitution $\sigma$ (both finite), denoted $\mathcal{D}om(\sigma)$ and $\mathcal{R}an(\sigma)$ respectively, as follows:
\begin{eqnarray*}
\mathcal{D}om(\sigma)&~=~&\{x \in V \mid \sigma(x) \neq x\}~,\\
\mathcal{R}an(\sigma)&~=~&\{t \in T_{\Sigma} \mid \exists x \cdot x \in \mathcal{D}om(\sigma) \land t = \sigma(x)\}~.
\end{eqnarray*}
Note that the application of a substitution $\sigma$ to a term $l$ \textit{simultaneously} replaces occurrences of variables by their respective $\sigma$-images. For the rest of this work, we restrict substitutions according to the following definition:
\begin{definition}[Non-conflicting Substitution]
A substitution $\sigma$ is said to be non-conflicting iff
\begin{eqnarray*}
[\bigcup_{t \in \mathcal{R}an(\sigma)}\mathcal{V}ar(t)] \cap \mathcal{D}om(\sigma)~=~\emptyset~.
\end{eqnarray*}
\end{definition}
Intuitively, a non-conflicting substitution can be simulated by a syntactic replacement as follows:
\begin{eqnarray*}
\sigma(l)~\defi~[x_1:=\sigma(x_1)]...[x_n:=\sigma(x_n)]l~.
\end{eqnarray*}
such that $x_1,...,x_n$ are the free variables in $l$, and $x_i \backslash \sigma(x_j)$ for all $i$ and $j$ where $1\leq i\leq n$ and $1\leq j\leq n$. In this case, we have the following important property:
\begin{eqnarray*}
\mathcal{D}(\sigma(l))~\leqv~\bigwedge_{e \in \mathcal{R}an(\sigma)}\mathcal{D}(e)~\land~\sigma(\mathcal{D}(l))~,
\end{eqnarray*}
which can proved by induction on the structure of terms.
\par
One of the main concepts of term rewriting is that of positions in terms and formulae where $\epsilon$ denotes the root position. Positions within a formula (or a term) describe paths to its subterms and subformulae. When $p$ is a position in a formula $F$, we write $F|_p$ for the term or formula at position $p$ in formula $F$. We write $F[s]_p$ for the formula that results from replacing $F|_p$ with $s$ in $F$. 
\section{WD-Preserving Rewriting}\label{wdpres}
In this section, we show how rewriting \textit{preserves} equality of terms, validity of formulae and well-definedness of both terms and formulae. The next definitions describe what is meant by a conditional rewrite rule.

\begin{definition}[Conditional Identity]
A $\Sigma$-conditional identity (simply conditional identity) is a triplet $(l, c, r)\in T_{\Sigma}\times F_{\Sigma}\times T_{\Sigma}$. In this case, $l$ is called the left hand side, $r$ the right hand side, and $c$ the condition of the identity.
\end{definition}

\begin{definition}[Valid Conditional Identity]\label{validCI}
A conditional identity $(l, c, r)$ is valid iff the following sequent is provable
\begin{eqnarray*}
c~\vdash_{_\mathcal{D}}~l=r~.
\end{eqnarray*}
\end{definition}

A conditional identity can be turned into a rewrite rule if it satisfies the syntactic restrictions presented in the following definition:

\begin{definition}[Conditional Term Rewrite Rule]
A conditional term rewrite rule is a conditional identity $(l, c, r)$ such that:
\begin{enumerate}
\item $l$ is not a variable,
\item $\mathcal{V}ar(c) \subseteq \mathcal{V}ar(l)$,
\item $\mathcal{V}ar(r) \subseteq \mathcal{V}ar(l)$.
\end{enumerate}
In this case, we use the notation $l \xrightarrow{c} r$ instead of $(l, c, r)$.
\end{definition}

In the derivations of Figure \ref{gRewDeriv} and Figure \ref{hRewDeriv}, we single out the necessary conditions under which rewriting can be performed. Figure \ref{hRewDeriv} concerns the rewriting of an hypothesis that has an occurrence of a rewrite rule left hand side $l$. Note the presence of the condition $\sigma(c)$. We assume that the free variables of $\sigma(c)$ also occur free in $\varphi[\sigma(l)]_p$; this ensures that $\sigma$ denotes the same substitution in both $\sigma(c)$ and $\varphi[\sigma(l)]_p$. Figure \ref{gRewDeriv}, on the other hand, concerns the rewriting of a goal which has an occurrence of a rewrite rule left hand side $l$.
\begin{figure}[h]
\centering
\fbox{
\begin{footnotesize}
$$
\infer[cut_{_\mathcal{D}}]
{\sigma(c), \varphi[\sigma(l)]_p~\vdash_{_\mathcal{D}}~R}
{
\infer[hyp_{_{WD}}]{\sigma(c), \varphi[\sigma(l)]_p~\vdash_{_\mathcal{D}}~\mathcal{D}(\varphi[\sigma(r)]_p)}{\boxed{\sigma(c), \mathcal{D}(\varphi[\sigma(l)]_p)~\vdash_{_\mathcal{D}}~\mathcal{D}(\varphi[\sigma(r)]_p)}}~~\boxed{\sigma(c), \varphi[\sigma(l)]_p~\vdash_{_\mathcal{D}}~\varphi[\sigma(r)]_p}~~
\infer[mon_{_{\mathcal{D}}}]{\sigma(c), \varphi[\sigma(l)]_p, \varphi[\sigma(r)]_p~\vdash_{_\mathcal{D}}~R}{\sigma(c), \varphi[\sigma(r)]_p~\vdash_{_\mathcal{D}}~R}
}
$$
\end{footnotesize}
}
\caption{Hypothesis Rewriting}
\label{hRewDeriv}
\end{figure}

\begin{figure}[h]
\centering
\fbox{
\begin{footnotesize}
$$
\infer[goal_{_{WD}}]
{\sigma(c)~\vdash_{_\mathcal{D}}~\varphi[\sigma(l)]_p}
{\infer[cut_{_\mathcal{D}}]{\sigma(c), \mathcal{D}(\varphi[\sigma(l)]_p)~\vdash_{_\mathcal{D}}~\varphi[\sigma(l)]_p}
	{\boxed{\sigma(c), \mathcal{D}(\varphi[\sigma(l)]_p)~\vdash_{_\mathcal{D}}~\mathcal{D}(\varphi[\sigma(r)]_p)}~~
	\infer[mon_{_{\mathcal{D}}}]{\sigma(c), \mathcal{D}(\varphi[\sigma(l)]_p)~\vdash_{_\mathcal{D}}~\varphi[\sigma(r)]_p}{\sigma(c)~\vdash_{_\mathcal{D}}~\varphi[\sigma(r)]_p}~~
	\boxed{\sigma(c), \varphi[\sigma(r)]_p~\vdash_{_\mathcal{D}}~\varphi[\sigma(l)]_p}}}
	
$$
\end{footnotesize}
}
\caption{Goal Rewriting}
\label{gRewDeriv}
\end{figure}

The boxed sequents correspond to the conditions under which a formula (an hypothesis or the goal) can be rewritten. In summary, a conditional term rewrite rule $l \xrightarrow{c} r$ can be applied to a formula $\varphi[\sigma(l)]_p$ (the goal or one of the hypothesises) iff the following sequents are provable:
\begin{eqnarray}
\sigma(c), \mathcal{D}(\varphi[\sigma(l)]_p)&~\vdash_{_{\mathcal{D}}}~&\mathcal{D}(\varphi[\sigma(r)]_p) \label{exCon1}\\
\sigma(c)&~\vdash_{_{\mathcal{D}}}~&\varphi[\sigma(l)]_p \leqv \varphi[\sigma(r)]_p\label{exCon2}~.
\end{eqnarray}

In the rest of this section, we examine the sufficient restrictions on conditional term rewrite rules to ensure that sequents \ref{exCon2} and \ref{exCon1} are provable for a given formula $\varphi$, a position $p$ and a substitution $\sigma$.

\begin{definition}\label{wdRewPres}
A conditional rewrite rule $l \xrightarrow{c} r$ is said to be WD-preserving iff the following sequent is provable:
\begin{eqnarray*}
\mathcal{D}(l), c~\vdash_{_\mathcal{D}}~\mathcal{D}(r)~.
\end{eqnarray*}
\end{definition}

We turn our attention to rewrite rule application. Consider applying rule $l \xrightarrow{c} r$ to formulae $P[s]_p$ where $s$ is a term as is $P|_p$. The left hand side $l$ is matched against $s$ by finding a substitution $\sigma$ such that $\sigma(l)=s$ (one-way matching). Provided $\sigma(c)$ holds,  $P[s]_p$ can be rewritten to $P[\sigma(r)]_p$. 
\par
The following theorem states that the application of a valid and well-definedness preserving conditional term rewrite rule preserves equality (\ref{termVal}) and well-definedness (\ref{termWD}) of terms.
\begin{theorem}\label{thm1}
Let $l \xrightarrow{c} r$ be a conditional term rewrite rule, $t$ be a term, $p$ be a position within $t$, and $\sigma$ be a non-conflicting substitution such that
\begin{eqnarray*}
\mathcal{D}om(\sigma)\subseteq \mathcal{V}ar(l)~.
\end{eqnarray*}
If $l \xrightarrow{c} r$ is valid and WD-preserving, then the following two sequents are provable:
\begin{eqnarray}
\sigma(c)&~\vdash_{_\mathcal{D}}~&t[\sigma(l)]_p = t[\sigma(r)]_p~, \label{termVal}\\
\mathcal{D}(t[\sigma(l)]_p), \sigma(c)&~\vdash_{_\mathcal{D}}~&\mathcal{D}(t[\sigma(r)]_p)~\label{termWD}.
\end{eqnarray}
\end{theorem}
\noindent \textbf{\textit{Proof}. } The following lemma is needed to prove Theorem \ref{thm1}:
\begin{lemma}\label{lemma1}
Let $l \xrightarrow{c} r$ be a conditional term rewrite rule, and $\sigma$ be a non-conflicting substitution such that
\begin{eqnarray*}
\mathcal{D}om(\sigma) \subseteq \mathcal{V}ar(l)~.
\end{eqnarray*}
\begin{enumerate}
\item If $l \xrightarrow{c} r$ is valid, then the following sequent is provable:
\begin{eqnarray*}
\sigma(c)~\vdash_{_\mathcal{D}}~\sigma(l)=\sigma(r)~.
\end{eqnarray*}
\item If $l \xrightarrow{c} r$ is WD-preserving, then the following sequents are provable:
\begin{eqnarray*}
\mathcal{D}(\sigma(l)) \land \sigma(c)&~\vdash_{_\mathcal{D}}~&\mathcal{D}(\sigma(r))~.
\end{eqnarray*}
\end{enumerate}
\end{lemma}
\noindent \textbf{\textit{Proof}. }
We observe that the sequent
\begin{eqnarray}
\vdash_{_\mathcal{D}}~\forall \overrightarrow{x}\cdot [(\mathcal{D}(l)\land \mathcal{D}(c) \land \mathcal{D}(r)\land c)\limp l=r]\label{lemma1seq1}
\end{eqnarray}
($\overrightarrow{x}$ are the free variables of $l$) is provable if the sequent
\begin{eqnarray*}
c~\vdash_{_\mathcal{D}}~l=r
\end{eqnarray*}
is also provable. We also observe that the sequent
\begin{eqnarray}
\vdash_{_\mathcal{D}}~\mathcal{D}(\forall \overrightarrow{x}\cdot [(\mathcal{D}(l)\land \mathcal{D}(c) \land \mathcal{D}(r)\land c)\limp l=r])
\end{eqnarray}
is provable. Since the substitution $\sigma$ can be simulated as a sequence of syntactic replacements, instantiating $\overrightarrow{x}$ in (\ref{lemma1seq1}) with the appropriate terms in $\mathcal{R}an(\sigma)$ is the main idea of the proof of the first claim. The proof of the second claim follows a similar approach.

\begin{enumerate}
\item \textit{Proof of sequent} (\ref{termVal}): We proceed by induction on the structure of the term $t$.
\begin{enumerate}
\item \textbf{Base Case:} $t$ is a variable, $t = x$. In this case (\ref{termVal}) becomes
\begin{eqnarray*}
\sigma(c)~\vdash_{_\mathcal{D}}~x[\sigma(l)]_\epsilon = x[\sigma(r)]_\epsilon~,
\end{eqnarray*}
since variables have only one position ($\epsilon$ the root position). This simplifies to 
\begin{eqnarray*}
\sigma(c)~\vdash_{_\mathcal{D}}~\sigma(l) = \sigma(r)~,
\end{eqnarray*}
which is a provable sequent according to Lemma \ref{lemma1}.
\item \textbf{Inductive Case:} $t$ is a function, $t = f(t_1,...,t_n)$. We distinguish the cases $p=\epsilon$ and $p=iq$ for $1\leq i\leq n$ and some position $q$.
\begin{enumerate}
\item Case $p=\epsilon$: this case is similar to the base case.
\item Case $p=iq$: we assume the following inductive hypothesis (in this case a provable sequent)
\begin{eqnarray*}
\sigma(c)~\vdash_{_\mathcal{D}}~t_i[\sigma(l)]_q = t_i[\sigma(r)]_q~,
\end{eqnarray*}
and we show that
\begin{small}
\begin{eqnarray*}
\sigma(c)~\vdash_{_\mathcal{D}}~f(t_1,...,t_i[\sigma(l)]_q,...,t_n) = f(t_1,...,t_i[\sigma(r)]_q,...,t_n)~,
\end{eqnarray*}
\end{small}
is a provable sequent where $iq=p$.
\end{enumerate}
\end{enumerate}
\item \textit{Proof of sequent} (\ref{termWD}): We proceed by induction on the structure of the term $t$.
\begin{enumerate}
\item \textbf{Base Case:} $t$ is a variable, $t = x$. In this case (\ref{termWD}) becomes
\begin{eqnarray*}
\mathcal{D}(x[\sigma(l)]_\epsilon), \sigma(c)~\vdash_{_\mathcal{D}}~\mathcal{D}( x[\sigma(r)]_\epsilon)~,
\end{eqnarray*}
since variables only have the root position $\epsilon$. This simplifies to 
\begin{eqnarray*}
\mathcal{D}(\sigma(l)), \sigma(c)~\vdash_{_\mathcal{D}}~\mathcal{D}(\sigma(r))~,
\end{eqnarray*}
which is a provable sequent according to Lemma \ref{lemma1}.
\item \textbf{Inductive Case:} $t$ is a function, $t = f(t_1,...,t_n)$. We distinguish the cases $p=\epsilon$ and $p=iq$ for $1\leq i\leq n$ and some position $q$.
\begin{enumerate}
\item Case $p=\epsilon$: this case is similar to the base case.
\item Case $p=iq$: We assume the following inductive hypothesis
\begin{eqnarray*}
\mathcal{D}(t_i[\sigma(l)]_q), \sigma(c)~\vdash_{_\mathcal{D}}~\mathcal{D}(t_i[\sigma(r)]_q)~,
\end{eqnarray*}
and we show that
\begin{small}
\begin{eqnarray*}
\mathcal{D}(f(t_1,...,t_i[\sigma(l)]_q,...,t_n)), \sigma(c)~\vdash_{_\mathcal{D}}~
\mathcal{D}(f(t_1,...,t_i[\sigma(r)]_q,...,t_n))~,
\end{eqnarray*}
\end{small}
is a provable sequent where $iq=p$.
\end{enumerate}
\end{enumerate}
\end{enumerate}

The following theorem asserts that Definition \ref{validCI} and \ref{wdRewPres} are adequate for a conditional term rewrite rule to preserve validity and well-definedness when applied to a formula.
\begin{theorem}\label{thm2}
Let $l \xrightarrow{c} r$ be a conditional term rewrite rule, $f$ be a formula, $p$ be a position within $f$ such that $f|_P$ is a term, and $\sigma$ be a non-conflicting substitution such that
\begin{eqnarray*}
\mathcal{D}om(\sigma) \subseteq \mathcal{V}ar(l)~.
\end{eqnarray*}
If $l \xrightarrow{c} r$ is valid and WD-preserving, then the following two sequents are provable:
\begin{eqnarray}
\sigma(c)&~\vdash_{_\mathcal{D}}~&f[\sigma(l)]_p \leqv f[\sigma(r)]_p~, \label{formulaVal}\\
\mathcal{D}(f[\sigma(l)]_p), \sigma(c)&~\vdash_{_\mathcal{D}}~&\mathcal{D}(f[\sigma(r)]_p)~\label{formulaWD}.
\end{eqnarray}
\end{theorem}
\noindent \textbf{\textit{Proof}. }
\begin{enumerate}
\item \textit{Proof of sequent} (\ref{formulaVal}): We proceed by induction on the structure of the formula $f$. We show a sketch of the proof, and only cover three interesting cases.
\begin{enumerate}
\item \textbf{Base Case:} $f$ is of the shape $r(t_1,...,t_n)$ such that $r \in P$ and $t_1,...,t_n$ are terms. In this case, position $p$ can only be of the form $iq$ for some position $q$ and $1\leq i \leq n$ since the root position is of a formula.
Therefore, (\ref{formulaVal}) becomes
\begin{small}
\begin{eqnarray*}
\sigma(c)~\vdash_{_\mathcal{D}}~r(t_1,...,t_n)[\sigma(l)]_p \leqv r(t_1,...,t_n)[\sigma(r)]_p~,
\end{eqnarray*}
\end{small}
where $p=iq$ for some position $q$ and $1\leq i \leq n$. This can be rewritten to
\begin{small}
\begin{eqnarray*}
\sigma(c)~\vdash_{_\mathcal{D}}~r(t_1,...,t_i[\sigma(l)]_q,...,t_n) \leqv r(t_1,...,t_i[\sigma(r)]_q,...,t_n)~.
\end{eqnarray*}
\end{small}
This amounts to proving the following two sequents:
\begin{small}
\begin{eqnarray*}
\sigma(c), r(t_1,...,t_i[\sigma(l)]_q,...,t_n)~\vdash_{_\mathcal{D}}~r(t_1,...,t_i[\sigma(r)]_q,...,t_n)~,\\
\sigma(c), r(t_1,...,t_i[\sigma(r)]_q,...,t_n)~\vdash_{_\mathcal{D}}~r(t_1,...,t_i[\sigma(l)]_q,...,t_n)~.
\end{eqnarray*}
\end{small}
Using Theorem \ref{thm1}, both sequents can be shown to be provable.
\item \textbf{Inductive Case:} $f$ is of the shape $\varphi \land \psi$ such that $\varphi$ and $\psi$ are formulae. In this case, (\ref{formulaVal}) becomes
\begin{small}
\begin{equation}
\sigma(c)~\vdash_{_\mathcal{D}}~(\varphi \land \psi)[\sigma(l)]_p \leqv (\varphi \land \psi)[\sigma(r)]_p~.\label{seq3.5}
\end{equation}
\end{small}
Position $p$ can only be of the form $p=1q$ or $p=2q$ for some position $q$. We distinguish the two cases:
\begin{enumerate}
\item $p=1q$: In this case, sequent (\ref{seq3.5}) becomes
\begin{small}
\begin{equation}
\sigma(c)~\vdash_{_\mathcal{D}}~(\varphi[\sigma(l)]_q \land \psi) \leqv (\varphi[\sigma(r)]_q \land \psi)~.\label{seq3.6}
\end{equation}
\end{small}
To proceed, we assume the following inductive hypothesis
\begin{small}
\begin{equation}
\sigma(c)~\vdash_{_\mathcal{D}}~(\varphi[\sigma(l)]_q) \leqv (\varphi[\sigma(r)]_q)~,\label{seq3.7}
\end{equation}
\end{small}
and we show that sequent (\ref{seq3.6}) is provable.
\item $p=2q$: analogous to the previous case.
\end{enumerate}
\item \textbf{Inductive Case:} $f$ is of the shape $\forall x\cdot\varphi$ such that $\varphi$ is a formula. In this case, (\ref{formulaVal}) becomes
\begin{small}
\begin{equation}
\sigma(c)~\vdash_{_\mathcal{D}}~(\forall x \cdot \varphi)[\sigma(l)]_p \leqv (\forall x \cdot \varphi)[\sigma(r)]_p~.\label{seq3.8}
\end{equation}
\end{small}
Position $p$ can only be of the form $p=1q$ for some position $q$. Sequent (\ref{seq3.8}) simplifies to
\begin{small}
\begin{equation}
\sigma(c)~\vdash_{_\mathcal{D}}~(\forall x \cdot \varphi[\sigma(l)]_q) \leqv (\forall x \cdot \varphi[\sigma(r)]_q)~.\label{seq3.9}
\end{equation}
\end{small}
To proceed, we assume that the following sequent is provable:
\begin{small}
\begin{equation}
\sigma(c)~\vdash_{_\mathcal{D}}~(\varphi[\sigma(l)]_q) \leqv (\varphi[\sigma(r)]_q)~,\label{seq3.10}
\end{equation}
\end{small}
and we show that sequent (\ref{seq3.9}) is provable.
\end{enumerate}
\item \textit{Proof of sequent} (\ref{formulaWD}):
is similar to the proof of sequent (\ref{formulaVal}). We only show one inductive case.
\begin{enumerate}
\item \textbf{Inductive Case:} $f$ is of the shape $\varphi \land \psi$ such that $\varphi$ and $\psi$ are formulae. In this case, (\ref{formulaWD}) becomes
\begin{small}
\begin{equation}
\mathcal{D}((\varphi \land \psi)[\sigma(l)]_p), \sigma(c)~\vdash_{_\mathcal{D}}~\mathcal{D}((\varphi \land \psi)[\sigma(r)]_p)~.\label{seq3.11}
\end{equation}
\end{small}
Position $p$ can only be of the form $p=1q$ or $p=2q$ for some position $q$. We distinguish the two cases:
\begin{enumerate}
\item $p=1q$: In this case, sequent (\ref{seq3.11}) becomes
\begin{small}
\begin{equation}
\mathcal{D}((\varphi[\sigma(l)]_q \land \psi)), \sigma(c)~\vdash_{_\mathcal{D}}~\mathcal{D}((\varphi[\sigma(r)]_q \land \psi))~.\label{seq3.12}
\end{equation}
\end{small}
To proceed, we assume that the following sequent is provable:
\begin{small}
\begin{equation}
\mathcal{D}((\varphi[\sigma(l)]_q)), \sigma(c)~\vdash_{_\mathcal{D}}~\mathcal{D}((\varphi[\sigma(r)]_q))~,\label{seq3.13}
\end{equation}
\end{small}
and we show that sequent (\ref{seq3.12}) is provable.
\item $p=2q$: analogous to the previous case.
\end{enumerate}
\end{enumerate}
\end{enumerate}

\textbf{Summary.} In this section, we have defined the criteria for the validity and well-definedness preservation of term rewrite rules when rewriting interleaves with the rule of the proof system developed in~\cite{icfemMehta08}. In the next section, we show how rewriting can be systematically used as a proof step.
\section{Rewriting as a Proof Step}\label{pstep}
Rewriting can be used in proofs alongside the WD-preserving sequent calculus. Conditional term rewrite rules which have the same left hand side are grouped together. For this purpose, we use a more convenient notation. Given a valid and WD-preserving (grouped) conditional term rewrite rule
\begin{eqnarray*}
l \rightarrow &c_1: r_1\\
&...\\
&c_n: r_n
\end{eqnarray*}
we can add the following proof step to our calculus
\begin{eqnarray}
\frac
	{\left\{
	\begin{array}{l}
	H, P[\sigma(l)]_p~\vdash_{_\mathcal{D}}~\mathcal{D}(\sigma(c_1)\lor ...\lor \sigma(c_n)) \\
	H, P[\sigma(l)]_p~\vdash_{_\mathcal{D}}~\sigma(c_1)\lor ...\lor \sigma(c_n) \\
	H, \sigma(c_1), P[\sigma(r_1)]_p~\vdash_{_\mathcal{D}}~R~~...~~H, \sigma(c_n), P[\sigma(r_n)]_p~\vdash_{_\mathcal{D}}~R
	\end{array}\right.
	}
	{H, P[\sigma(l)]_p~\vdash_{_\mathcal{D}}~R}~\rightarrow hyp_{_\mathcal{D}}\label{hyp_app}
\end{eqnarray}
under the proviso that all free variables of $\sigma(r)$ (for all $i$ such that $1 \leq i \leq n$) occur free in $P[\sigma(l)]_p$. This proof step allows the hypothesis $P[\sigma(l)]_p$ to be rewritten to several cases according to the rewrite rule.
Under the proviso that all free variables of $\sigma(r_i)$ (for all $i$ such that $1 \leq i \leq n$) occur free in $R[\sigma(l)]_p$, the following proof step can be added for goal rewriting
\begin{eqnarray}
\frac
	{\left\{
	\begin{array}{l}
	H~\vdash_{_\mathcal{D}}~\mathcal{D}(\sigma(c_1)\lor ...\lor \sigma(c_n)) \\
	H~\vdash_{_\mathcal{D}}~\sigma(c_1)\lor ...\lor \sigma(c_n) \\
	H, \sigma(c_1)~\vdash_{_\mathcal{D}}~R[\sigma(r_1)]_p~~...~~H, \sigma(c_n)~\vdash_{_\mathcal{D}}~R[\sigma(r_n)]_p
	\end{array}\right.
	}
	{H~\vdash_{_\mathcal{D}}~R[\sigma(l)]_p}~\rightarrow goal_{_\mathcal{D}}~.\label{goal_app}
\end{eqnarray}
Proof steps (\ref{hyp_app}) and (\ref{goal_app}) can be derived using the cut rule, followed by a disjunction elimination after which rewriting can be applied. We now examine some special cases that can be used to facilitate proofs.
\subsection{Unconditional Term Rewrite Rules}
A term rewrite rule $l \xrightarrow{c} r$ is called \textit{unconditional} iff $c \defi \top$. In this case, steps (\ref{hyp_app}) and (\ref{goal_app}) can be simplified as follows:
\begin{eqnarray}
\frac
	{
	H, P[\sigma(r)]_p~\vdash_{_\mathcal{D}}~R
	}
	{H, P[\sigma(l)]_p~\vdash_{_\mathcal{D}}~R}~\rightarrow uhyp_{_\mathcal{D}}\label{hyp_app2}
\end{eqnarray}
\begin{eqnarray}
\frac
	{
	H~\vdash_{_\mathcal{D}}~R[\sigma(r)]_p
	}
	{H~\vdash_{_\mathcal{D}}~R[\sigma(l)]_p}~\rightarrow ugoal_{_\mathcal{D}}~.\label{goal_app2}
\end{eqnarray}
\subsection{Case-complete Grouped Term Rewrite Rules}
A grouped term rewrite rule 
\begin{eqnarray*}
l \rightarrow &c_1: r_1\\
&...\\
&c_n: r_n
\end{eqnarray*}
is called \textit{case-complete} iff the following sequent is provable:
\begin{eqnarray*}
\vdash_{_\mathcal{D}}~c_1\lor ...\lor c_n~.
\end{eqnarray*}
In this case, steps (\ref{hyp_app}) and (\ref{goal_app}) can be simplified as follows:
\begin{eqnarray}
\frac
	{\left\{
	\begin{array}{l}
	H, P[\sigma(l)]_p~\vdash_{_\mathcal{D}}~\mathcal{D}(\sigma(c_1)\lor ...\lor \sigma(c_n)) \\
	H, \sigma(c_1), P[\sigma(r_1)]_p~\vdash_{_\mathcal{D}}~R~~...~~H, \sigma(c_n), P[\sigma(r_n)]_p~\vdash_{_\mathcal{D}}~R
	\end{array}\right.
	}
	{H, P[\sigma(l)]_p~\vdash_{_\mathcal{D}}~R}~\rightarrow chyp_{_\mathcal{D}}\label{hyp_app3}
\end{eqnarray}
\begin{eqnarray}
\frac
	{\left\{
	\begin{array}{l}
	H~\vdash_{_\mathcal{D}}~\mathcal{D}(\sigma(c_1)\lor ...\lor \sigma(c_n)) \\
	H, \sigma(c_1)~\vdash_{_\mathcal{D}}~R[\sigma(r_1)]_p~~...~~H, \sigma(c_n)~\vdash_{_\mathcal{D}}~R[\sigma(r_n)]_p
	\end{array}\right.
	}
	{H~\vdash_{_\mathcal{D}}~R[\sigma(l)]_p}~\rightarrow cgoal_{_\mathcal{D}}~.\label{goal_app3}
\end{eqnarray}
\subsection{Top-level Occurrence}

\begin{definition}[Top-Level Occurrence]
Let $t$ be a term, $f$ be a formula, $p$ be a position within $f$. We say that $t$ has a top-level occurrence in $f$ if $f$ is either of the form 
\begin{enumerate}
\item $q(t_1,...,t_n)[t]_p$ where $q \in P$ and $t_1$,..., $t_n$ are terms, or; 
\item $(t_1=t_2)[t]_p$ where $t_1$ and $t_2$ are terms.  
\end{enumerate}
If $t$ has a top-level occurrence in $f$, then it also has a top-level occurrence in $\lnot f$.
\end{definition}
We have the following interesting property:
\begin{proposition}\label{toplevel_prop}
If the term $t$ has a top-level occurrence in formula $f$, then the following holds
\begin{eqnarray*}
\vdash_{_\mathcal{D}}~\mathcal{D}(f) \Rightarrow \mathcal{D}(t)~.
\end{eqnarray*}
\end{proposition}
If we further constrain grouped conditional term rewrite rules such that we have
\begin{eqnarray*}
\vdash_{_\mathcal{D}}~\mathcal{D}(l) \limp \bigwedge_{i=1}^{n}\mathcal{D}(c_i)~,
\end{eqnarray*}
Proposition \ref{toplevel_prop} can be used to simplify proofs. Let $P[\sigma(l)]_p$ be a formula such that $\sigma(l)$ occurs at the top-level. Since the grouped term rewrite rule is valid and WD-preserving, and using the previous proposition, we have the following
\begin{eqnarray*}
\vdash_{_\mathcal{D}}~\mathcal{D}(P[\sigma(l)]_p) \Rightarrow \mathcal{D}(\sigma(l))
\end{eqnarray*}
and, consequently:
\begin{eqnarray*}
\vdash_{_\mathcal{D}}~\mathcal{D}(P[\sigma(l)]_p) \Rightarrow \bigwedge_{i=1}^n \mathcal{D}(\sigma(c_i))
\end{eqnarray*}
under the proviso that all free variables of $\sigma(c_i)$ (for all $i$ such that $1 \leq i \leq n$) occur free in $P[\sigma(l)]_p$. In this particular case, the sequents
\begin{eqnarray*}
&H, P[\sigma(l)]_p~\vdash_{_\mathcal{D}}~\mathcal{D}(\sigma(c_1)\lor ...\lor \sigma(c_n))~,\\
&H, P~\vdash_{_\mathcal{D}}~\mathcal{D}(\sigma(c_1)\lor ...\lor \sigma(c_n))
\end{eqnarray*}
in (\ref{hyp_app}) and (\ref{goal_app}) respectively, are guaranteed to be provable. As such, they could be removed from the list of sub-goals that the modeller sees.
\section{Applications to Event-B}\label{eapp}
As mentioned in \ref{pracSetting}, Event-B modelling is carried out using two constructs: contexts and machines. A third construct, called \textit{theory}, has been implemented to bring a degree of meta-reasoning to the Rodin platform~\cite{3056425}. The theory construct has the following shape:
\begin{figure}[ht]
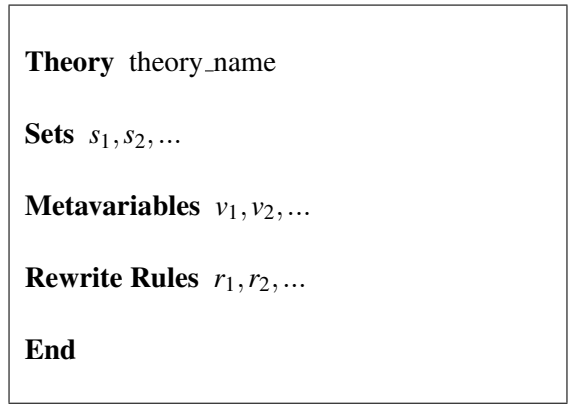

\begin{center}
\fbox{
\begin{minipage}{7cm}
\vspace{0.2in}
\textbf{Theory}~~theory\_name\\\\
\textbf{Sets}~~$s_1, s_2,...$\\\\
\textbf{Metavariables}~~$v_1, v_2,...$\\\\
\textbf{Rewrite Rules}~~$r_1, r_2,...$\\\\
\textbf{End}\\
\end{minipage}
}
\end{center}
\caption{The Theory Construct}\label{thy}
\end{figure}
\begin{enumerate}
\item \textit{Sets}. A theory can define a number of given sets which define the types on which the theory is parametrised.
\item \textit{Metavariables}. A theory can define a number of metavariables that can be used to specify rewrite rules. Each metavariable is associated with a type; this can be constructed using the given sets of the theory as well as the built-in types (e.g., $\mathbb{Z}$) using type constructors. For example, if a given set $S$ is defined within a theory, then $\pow{\mathbb{(Z)}}\times S$ can be used as a type for a metavariable.
\item \textit{Rewrite Rules}. Rewrite rules are one-directional equations that can be used to rewrite formulae to equivalent forms. As part of specifying a rewrite rule, the \textit{theory developer} decides whether the rule can be applied automatically without user intervention or interactively following a user request.
\end{enumerate}
The theory construct can be extended to enable the specification of inference rules. In brief, it facilitates the following:
\begin{itemize}
\item specification of proof rules within the same platform providing a degree of meta-reasoning within Rodin,
\item validation of specified proof rules to ensure that the soundness of the prover is not compromised.
\end{itemize}
The validation of rewrite rules is achieved by means of proof obligations. Definition \ref{validCI} and \ref{wdRewPres} defined the criteria for validity and WD-preservation of rewrite rules.
\par
The theory construct has been developed as part of a \textit{rule-based prover}~\cite{issam1984} which, in brief, offers the following capabilities:
\begin{enumerate}
\item Users can develop theories in the same way as contexts and machines. At the moment, theory development includes specification of rewrite rules including definition of sets and metavariables. Metavariables must be defined with their types which can be constructed from the theory sets and any built-in types (e.g., $\mathbb{Z}$) using type constructors (e.g., $\pow$).
\item Users can validate rewrite rules through generated proof obligations. The proof obligations generated for rules are to establish soundness, well-definedness preservation and case-completeness.
\item Users can deploy theories to a specific directory where they become available to the interactive and automatic provers of Rodin. Theory deployment adds soundness information to all deployed rules.
\item Users can use rewrite rules defined within the deployed theories as a part of the proving activity. A pattern matching mechanism is implemented to calculate applicable rewrite rules to any given sequent. 
\end{enumerate}
\vspace{2mm}
\textit{Examples.} The following two rules are valid and WD-preserving:
\begin{eqnarray}
card(i..j)~~&\xrightarrow{i \leq j}&~~j-i+1\\
card(i..j)~~&\xrightarrow{i > j}&~~0~,
\end{eqnarray}
where $i$ and $j$ are integers, $i..j$ denotes an integer range, and $card$ denotes the cardinality operator. The following rules are not WD-preserving:
\begin{eqnarray}
~~a&\xrightarrow{\top}&~~\frac{a}{a}\\
~~(f\ovl \{z\mapsto y\})(x)&\xrightarrow{x\neq z}&~~f(x)\label{unsRelO},
\end{eqnarray}
where $a$ is an integer, $f$ a relation, $x, y$, and $z$ are of arbitrary types. Moreover, `$\ovl$' denotes relational override. Rule \ref{unsRelO} is not WD-preserving since there could be a case where $f\ovl \{z\mapsto y\}$ is a function but $f$ is not. For instance, consider $f=\{1\mapsto 2, 1\mapsto 3, 2 \mapsto 4\}$, then 
$f\ovl\{1\mapsto 5\} = \{1\mapsto 5, 2 \mapsto 4\}$. In this case, $(f\ovl\{1\mapsto 5\})(1)$ is well-defined, but $f(1)$ is not.
\section{Future Work \& Conclusions}\label{future}
In this paper, we provided a treatment of well-definedness and rewriting. We singled out the necessary conditions under which rewriting preserves well-definedness. These conditions are necessary for the valid interleaving between rewriting steps and deduction in the WD-preserving proof calculus presented in~\cite{icfemMehta08}. In our study, we used the language signature $\Sigma$ whereby terms are only defined using other terms. In general, however, terms can also be constructed using formulae e.g., set comprehension $\{x\cdot P\}$. This changes the well-definedness conditions of terms, and it is interesting to establish whether the conditions outlined in Definition \ref{validCI} and \ref{wdRewPres} are indeed sufficient.
\par
We have presented a study unifying the notions of term rewriting and well-definedness in the context of the interleaving between deduction and rewriting. The results of this paper provided the theoretical foundations of an extensible rewriting-based prover (also called rule-based prover) that has been implemented for Event-B.

\bibliographystyle{plain}
\bibliography{par}

\begin{thebibliography}{10}

\bibitem{Abrial-etal06}
Jean-Raymond Abrial, Michael~J. Butler, Stefan Hallerstede, and Laurent Voisin.
\newblock {An Open Extensible Tool Environment for Event-B}.
\newblock In Zhiming Liu and Jifeng He, editors, {\em ICFEM}, volume 4260 of
  {\em Lecture Notes in Computer Science}, pages 588--605. Springer, 2006.

\bibitem{abrial_03_clickn}
Jean-Raymond Abrial and Dominique Cansell.
\newblock {Click'n Prove: Interactive Proofs within Set Theory}.
\newblock In David~A. Basin and Burkhart Wolff, editors, {\em TPHOLs}, volume
  2758 of {\em Lecture Notes in Computer Science}, pages 1--24. Springer, 2003.

\bibitem{1365974}
Jean-Raymond Abrial and Stefan Hallerstede.
\newblock {Refinement, Decomposition, and Instantiation of Discrete Models:
  Application to Event-B}.
\newblock {\em Fundam. Inform.}, 77(1-2):1--28, 2007.

\bibitem{723108}
Jean-Raymond Abrial and Louis Mussat.
\newblock {On Using Conditional Definitions in Formal Theories}.
\newblock In Didier Bert, Jonathan~P. Bowen, Martin~C. Henson, and Ken
  Robinson, editors, {\em ZB}, volume 2272 of {\em Lecture Notes in Computer
  Science}, pages 242--269. Springer, 2002.

\bibitem{280474}
Franz Baader and Tobias Nipkow.
\newblock {\em {Term Rewriting and All That}}.
\newblock Cambridge University Press, New York, NY, USA, 1998.

\bibitem{91938}
Ralph-Johan Back.
\newblock {Refinement Calculus, Part II: Parallel and Reactive Programs}.
\newblock In J.~W. de~Bakker, Willem~P. de~Roever, and Grzegorz Rozenberg,
  editors, {\em REX Workshop}, volume 430 of {\em Lecture Notes in Computer
  Science}, pages 67--93. Springer, 1989.

\bibitem{3056425}
Michael Butler and Stefan Hallerstede.
\newblock {The Rodin Formal Modelling Tool}.
\newblock {\em BCS-FACS Christmas 2007 Meeting - Formal Methods In Industry,
  London.}, December 2007.

\bibitem{303448}
Nachum Dershowitz.
\newblock {Book review: Term Rewriting Systems by "Terese" (Marc Bezem, Jan
  Willem Klop, and Roel de Vrijer, eds.), Cambridge University Press, Cambridge
  Tracts in Theoretical Computer Science 55, 2003, hard cover: ISBN
  0-521-39115-6}.
\newblock {\em TPLP}, 5(3):395--399, 2005.

\bibitem{949773}
Gilles Dowek, Th{\'e}r{\`e}se Hardin, and Claude Kirchner.
\newblock {Theorem Proving Modulo}.
\newblock {\em J. Autom. Reasoning}, 31(1):33--72, 2003.

\bibitem{LeuschelButler:FME03}
Michael Leuschel and Michael Butler.
\newblock {ProB: A Model Checker for B}.
\newblock In {\em FME 2003: Formal Methods, LNCS 2805}, pages 855--874.
  Springer-Verlag, 2003.

\bibitem{issam1984}
Issam Maamria, Michael Butler, Andrew Edmunds, and Abdolbaghi Rezazadeh.
\newblock {On an Extensible Rule-Based Prover for Event-B}.
\newblock In Marc Frappier, Uwe Gl{\"a}sser, Sarfraz Khurshid, R{\'e}gine
  Laleau, and Steve Reeves, editors, {\em ASM}, volume 5977 of {\em Lecture
  Notes in Computer Science}, page 407. Springer, 2010.

\bibitem{icfemMehta08}
Farhad Mehta.
\newblock {A Practical Approach to Partiality --- A Proof Based Approach}.
\newblock In {\em Proceedings of the 10th International Conference on Formal
  Methods and Software Engineering}, ICFEM '08, pages 238--257, Berlin,
  Heidelberg, 2008. Springer-Verlag.

\bibitem{farhad}
Farhad Mehta.
\newblock {\em Proofs for the Working Engineer}.
\newblock {PhD T}hesis, ETH Zurich, 2008.

\bibitem{513715}
Tobias Nipkow, Lawrence~C. Paulson, and Markus Wenzel.
\newblock {\em {Isabelle/HOL --- A Proof Assistant for Higher-Order Logic}},
  volume 2283 of {\em LNCS}.
\newblock Springer, 2002.

\bibitem{886667}
Sam Owre, John~M. Rushby, and Natarajan Shankar.
\newblock {PVS: A Prototype Verification System}.
\newblock In Deepak Kapur, editor, {\em CADE}, volume 607 of {\em Lecture Notes
  in Computer Science}, pages 748--752. Springer, 1992.

\end{thebibliography}
\end{document}